\documentclass [12pt]{article}
\usepackage{amsmath, amsthm, amssymb}
\usepackage{graphicx}
\usepackage{fullpage}
\usepackage{authblk}
\usepackage{mathrsfs}
\usepackage{bbm}
\usepackage{makecell}
\usepackage{bibentry, natbib}
\usepackage{float}
\usepackage[colorlinks, citecolor=blue]{hyperref}
\usepackage{tabularx}
\usepackage{booktabs}
\makeatletter 
\@addtoreset{equation}{section}
\makeatother  

\newtheorem{Theorem}{Theorem}
\newtheorem{Lemma}{Lemma}

\newtheorem{Definition}{Definition}
\newtheorem{Proposition}{Proposition}
\newtheorem{Example}{Example}
\usepackage{tikz}
\usepackage{pgflibraryarrows}
\usepackage{pgflibrarysnakes}

\begin{document}
\title{New axioms for top trading cycles\thanks{We are grateful to Zhenhua Jiao, Bettina Klaus, Peng Liu, Qianfeng Tang, and participants at the 2020 Interdisciplinary Forum of Game Theory and Economic Management for helpful comments and discussions. Yajing Chen's research is supported by The National Natural Science Foundation of China (No. 71703038), The Ministry of Education Project of Youth Fund of Humanities and Social Sciences (No. 17YJC790012), The Innovation Program of Shanghai Municipal Education Commission (No. 2017-01-07-00-02-E00008).Siwei Chen's research is supported by the National Natural Science Foundation of China (No. 71801230) and the Fundamental Research Funds for the Central Universities (No. 20wkpy43).}}
\author{Siwei Chen\footnote{Lingnan College, Sun Yat-sen University, Guangzhou, China
(email: chensw6@mail.sysu.edu.cn)}~~~~Yajing Chen\footnote{School of Business, East China University of Science and Technology, Shanghai, China. (Email: yajingchen@ecust.edu.cn; yajingchen87@gmail.com)}~~~~Chia-Ling Hsu\footnote{School of Economics, Southwestern University of Finance and Economics, Chengdu, China (Email: chialinghsu@hotmail.com; chialinghsu1982@gmail.com)}}
\date{\emph{First version: January 2013}\\ \emph{This version: April 2021}}
\maketitle

\begin{abstract}

School choice is of great importance both in theory and practice. This paper studies the (student-optimal) top trading cycles mechanism (TTCM) in an axiomatic way. We introduce two new axioms: MBG (mutual best group)-quota-rationality and MBG-robust efficiency. While stability implies MBG-quota-rationality, MBG-robust efficiency is weaker than robust efficiency, which is stronger than the combination of efficiency and group strategy-proofness. The TTCM is characterized by MBG-quota-rationality and MBG-robust efficiency. Our results construct a new basis to compare the TTCM with the other school choice mechanisms, especially the student-optimal stable mechanism under Ergin  but not Kesten-acyclic priority structures. 


\end{abstract}

\section{Introduction}

School choice (The priority-based allocation problem) considers the problem of allocating a set of school seats (each school may have multiple seats) to students on the basis of priorities, with each student being entitled to receive at most one seat. Monetary transfers are not allowed. In the seminal paper, \cite{abdulkadirouglu2003school} point out that the widely used Boston mechanism (BOSM) determined by the student-proposing immediate acceptance algorithm has serious shortcomings and suggest to replace it with the student-optimal stable mechanism (SOSM) determined by the student-proposing deferred acceptance algorithm (\cite{gale1962college}), or the top trading cycles mechanism (TTCM) determined by the student-proposing top trading cycles algorithm (\cite{shapley1974cores}). The above paper has inspired a lot of literature studying the school choice problem and influenced the practice of school choice deeply. However, at present, several important questions regarding TTCM still remain completely or partially unsolved.

The first question is how to characterize the TTCM when schools have multiple seats. Characterizing school choice mechanisms has attracted much attention recently. The SOSM has been characterized by \cite{morrill2013alternative} and \cite{chen2017new} on full acceptant priorities, while \cite{ehlers2014strategy} characterize it for all responsive priorities. The BOSM is characterized by \cite{afacan2013alternative}, and \cite{chen2016new}, respectively. Prior to our paper, other papers have also characterized the TTCM. \cite{abdulkadiroglu2010role} characterize it for the first time, by strategy-proofness, efficiency, and recursive respect of top priorities. \cite{morrill2011alternative} characterizes the TTCM by strategy-proofness or weak Maskin monotonicity, together with efficiency, mutual best, and independence of irrelevant rankings. \cite{abdulkadiroglu2010role} and \cite{morrill2011alternative} both get their results when each school has only one seat available. \cite{kesten2006two} shows the conditions under which the TTCM recovers resource monotonicity, population monotonicity, consistency, and is equivalent to the SOSM. \cite{abdulkadirouglu2020efficiency} finds that the TTCM is justified envy minimal among all Pareto efficient and strategy-proof mechanisms in one-to-one matching.

As stated above, appealing characterizations of the TTCM when schools have multiple seats is vacant. This paper aims to fill in this vacancy. We first define a key concept called  $``$mutual best group$"$, which reflects the phenomenon of spontaneous exchanges within people as widely observed in practice. A subset of students constitute a \emph{mutual best group}\footnote{This concept is similar to ``top fair set'' proposed by \cite{rong2020stable}.} (Definition 1) of a problem if it is the largest subset satisfying that the set of students with the highest priority for members of the set of favorite schools of students in this subset is equivalent to the original subset. Consequently, we can partition the set of students into a sequence of mutual best groups, and define a sequence of subproblems. We then introduce two new axioms based on ``mutual best group'': MBG (mutual best group)-quota-rationality and MBG-robust efficiency. The former is a fairness requirement which is implied by stability. The latter is a combination of efficiency and immunity to group manipulations. The TTCM is characterized by MBG-quota-rationality and MBG-robust efficiency. This characterization result constructs a new basis to compare the TTCM with the other school choice mechanisms.

The second question is what is the difference between TTCM and SOSM when the priority structure is Ergin-acyclic. In the insightful paper of \cite{ergin2002efficient}, the author shows that SOSM recovers Pareto efficiency, group strategy-proofness and consistency if and only the priority structure is Ergin-acyclic. However, SOSM is not equivalent to TTCM under Ergin-acyclic priority structures. \cite{kesten2006two} later shows that these two are equivalent, or TTCM recovers fairness, resource monotonicity and population monotonicity if and only if the priority structure is Kesten-acyclic. Ergin-acyclicity is weaker than Kesten-acyclicity since the former exclude less cycles.\footnote{See \cite{kesten2006two}, \cite{Haeringer2009Constrained}, and \cite{chen2021acyclic}.} Under Ergin-acyclic priority structures, when SOSM recovers desirable properties yet is not equivalent to TTCM, a natural question arises as what distinguishes TTCM from SOSM under such circumstances. We find that in the domain of Ergin-acyclic priority profiles, the TTCM satisfies MBG-collusion-proofness while the SOSM violates this axiom. It suggests that immunity to group manipulation and reallocation is the key advantage of TTCM over SOSM. 

To define MBG-robust efficiency, we need to first define robust efficiency. A mechanism satisfies \emph{robust efficiency}\footnote{This paper is not the first discussing axioms with robust flavor in school choice setting. \cite{kojima2011robust} and \cite{afacan2012Group} define robust stability and group stability, respectively.} if (1) it is Pareto efficient; and (2) it is collusion-proof, i.e., no subset of students can gain (all members are weakly better off while some members are strictly better off) by either reporting a false preference profile or first reporting a false preference profile and then reallocate their assigned schools within the subset. Building on the definition of mutual best group and robust efficiency, we can now define MBG-robust efficiency. A mechanism satisfies \emph{MBG-robust efficiency} if (1) it is Pareto efficient; and (2) it is MBG-collusion-proof, i.e., no mutual best group can gain (all members are weakly better off while some members are strictly better off) by either reporting a false preference profile or first reporting a false preference profile and then reallocate their assigned schools within the group. The TTCM satisfies MBG-robust efficiency but the SOSM violates it under Ergin-acyclic priority structures, as shown in example 2. This implies that TTCM outperforms SOSM under Ergin-acyclic priority structures in that it can prevent at least students within any mutual best group from colluding.

This paper is organized as follows: Section~2 introduces the model and the TTCM. Section~3 defines the key concept, mutual best group. Sections 4 introduces two new axioms related to mutual best group. We present the main characterization result in Section 5, and conclude this paper with some discussions in Section 6.

\section{Preliminaries}
\subsection{Notations}

Let $I$ denote a finite set of \textbf{students} and $O$ denote a finite set of \textbf{schools}. Each school $a\in O$ has a capacity $q_a$ with $q_a\in \mathbb{Z}_{++}$, which represents the number of seats available for school $a$. Let $q=(q_a)_{a\in O}$ be the \textbf{capacity vector} of all schools $O$. There is a null school $\varnothing$, and $q_{\varnothing}=\infty$. Let $\widetilde{O}= O \cup \{\varnothing\}$. Let $\mathbb{O}$ be the set of school seats, where seats in each school are treated as distinct objects.

A \textbf{matching} is a function $\mu: I \rightarrow \widetilde{O}$ where $( \romannumeral 1)$ for each $i\in I$, $\mu(i) \in \widetilde{O} $; and $( \romannumeral 2)$ for each $a\in O$, $|\mu^{-1}(a)|\leq q_a$. To simplify the notation, let $\mu _i =\mu(i)$ and $\mu _a =\mu^{-1}(a)$. For each $I' \subset I$, let $\mu_{I'} \equiv \{\mu_{i}|{i \in I'}\}$ be the set of school seats assigned to $I'$ under $\mu$. For each $I'\subset I$, let $\mu_{-I'}$ be the \textbf{reduced matching} by removing $I'$ and $\mu_{I'}$, and keeping the assignments of the other students unchanged. For each $I' \subseteq I$ and $\mathbb{O}'\subset \mathbb{O}$ such that $|I'|=|\mathbb{O}'|$, let $M(I', \mathbb{O}')$ be the set of all possible matchings (bijections) between $I'$ and $\mathbb{O}'$, and $m(I', \mathbb{O}')$ be a generic element of $M(I', \mathbb{O}')$.

Each student $i\in I$ has a strict preference order $P_i$ over $\widetilde{O}$. Let $R_i$ denote the corresponding weak preference of $P_i$, that is, $a R_i b $ if and only if $a P_i b$ or $a = b$.  A school $a$ is \textbf{acceptable} to a student $i$ if $a P_i \varnothing$. Denote by $\mathcal{P}$ the set of all such orders. Let $P=(P_i)_{i\in I}$ denote the \textbf{preference profile} of all students. Let $P_{I'}=(P_i)_{i\in I'}$ denote the preferences of any subset $I' \subset I$.

Each school $a \in O$ has a strict priority order $\succ_a$ over $I$, whereas $i \succ_a j$ means that student $i$ has higher priority than student $j$ at school $a$.  The null school $\varnothing$ is equipped with an exogenously given priority order $\succ_{\varnothing}$.  Let $\succ =(\succ_a)_{a\in \widetilde{O}}$ denote the \textbf{priority profile} of all schools.

To sum up, a school choice \textbf{problem} is a five-tuple $\epsilon \equiv (I, O, P, \succ, q)$. For simplicity of notations, we denote a problem by the preference profile $P$ and only denote it by $\epsilon$ when necessary, and these two are sometimes used exchangeably. Let $\mathcal{P}^{|I|}$, $\mathcal{M}$, and $\mathcal{E}$ be the set of all preference profiles, matchings and problems, respectively.

A school choice \textbf{mechanism} $\varphi$ finds a matching for each problem (preference profile). At $\epsilon$, student $i$ is assigned to $\varphi_i(\epsilon)$, and school $a$ is assigned to the set of students $\varphi_a(\epsilon)$. For each $I' \subseteq I$, let $\varphi_{I'}(\epsilon)$ be the set of school seats assigned to $I'$.

A matching $\nu$ Pareto dominates another matching $\mu$ at a problem $P$ if $( \romannumeral 1)$ $\nu_i R_i \mu_i$ for all $i \in I$, and $( \romannumeral 2)$ $\nu_i P_i \mu_i$ for some $i \in I$. A matching $\mu \in \mathcal{M}$ is \textbf{Pareto efficient} if it is not Pareto dominated by any other matching $\nu\in \mathcal{M}$. A mechanism $\varphi$ is \textbf{Pareto efficient} if it assigns a Pareto efficient matching to each problem. A matching $\mu \in \mathcal{M}$ is \textbf{non-wasteful} if for each $i\in I$ and $a \in \widetilde{O}$, $a P_i \mu_i$ implies $|\mu_a|=q_a$, and it is \textbf{fair} if for each $i\in I$ and $a \in O$, $a P_i \mu_i$ implies $j \succ_a i$ for each $j\in \mu_a$. A matching $\mu \in \mathcal{M}$ is \textbf{stable} if it is non-wasteful and fair. A mechanism $\varphi$ is \textbf{stable} if it assigns a stable matching to each problem.

A mechanism $\varphi$ is \textbf{strategy-proof} if truthful revelation of preferences is a weakly dominant strategy for each student, i.e., for each problem $P$, there exists no $i\in I$ and $P_i^{'} \in \mathcal{P}$ such that $\varphi_i(P_i^{'}, P_{-i})P_i \varphi_i(P)$. A mechanism $\varphi$ is \textbf{group strategy-proof} if for each problem $P$, there exists no nonempty $I^{'}\subseteq I$ and $P_{I^{'}}^{'} \in \mathcal{P}^{|I^{'}|}$ such that $( \romannumeral 1)$ $\varphi_i(P^{'}_{I^{'}}, P_{-I^{'}})  R_i \varphi_i(P)$ for all $i\in I^{'}$, and $( \romannumeral 2)$ $\varphi_i(P^{'}_{I^{'}}, P_{-I^{'}})  P_i \varphi_i(P)$ for some $i\in I^{'}$. A mechanism $\varphi$ is \textbf{reallocation-proof} if there exists no $I' \subseteq I$ where $|I'|=2$, and $P_{I^{'}}^{'} \in \mathcal{P}^2$ such that $( \romannumeral 1)$ $\varphi_i(P'_{i} ,P_{-i}) = \varphi_i(P)\neq \varphi_i(P'_{I'} ,P_{-I'})$ for each $i\in I'$, $( \romannumeral 2)$ $\varphi_j(P'_{I'} ,P_{-I'})R_i \varphi_i(P)$ for all $i,j\in I'$, and $( \romannumeral 3)$ $\varphi_j(P'_{I'} ,P_{-I'})P_i \varphi_i(P)$ for some $i,j\in I'$.\footnote{See Definition 5 of \cite{papai2000strategyproof} for more details. \cite{2018endowments} also define a notion of reallocation-proofness in section 4.3 of their paper. But the meaning of their axiom is different from \cite{papai2000strategyproof}'s. For consistency of terminology, we follow the definition in \cite{papai2000strategyproof}.}

\subsection{The (student-optimal) top trading cycles algorithm}

The top trading cycles school choice mechanism, denoted by $\varphi^T$, finds a matching via the following student-proposing \textbf{top trading cycles algorithm}:

\textbf{Step 1:} Each student points to his favorite school and each school points to the student who has the highest priority for it. There is at least one cycle. Each student in a cycle is assigned the school he points to and is removed. The capacities of schools are renewed accordingly and schools with zero capacity left are also removed.

\vdots

\textbf{Step} {\boldmath $k, k\geq 2$:} Each remaining student points to his favorite school among the remaining ones and each remaining school points to the student who has the highest priority for it among the remaining students. There is at least one cycle. Each student in a cycle is assigned the school he points to and is removed. The capacities of schools are renewed accordingly and schools with zero capacity left are also removed.
\newline
\newline
\indent The algorithm stops when all students have been removed. The top trading cycles algorithm is proposed in \cite{shapley1974cores} and later generalized by \cite{papai2000strategyproof} and \cite{pycia2017incentive}. In school choice setting, the TTCM based on this algorithm satisfies Pareto efficiency and group strategy-proofness but violates stability (\cite{abdulkadirouglu2003school}).

\section{Mutual best group}

\subsection{Definition}
Given a problem $\epsilon$, a subset of students $I' \subseteq I$ and a subset of schools $O^{'} \subseteq \widetilde{O}$, let
\begin{center}
$B(\epsilon, I') = \{a \in \widetilde{O} | \exists i \in I', aP_{i}b, \forall b \in \widetilde{O}  \setminus \{a\}\}$

$T(\epsilon, O')= \{i \in I | \exists a \in O', i \succ_{a} j , \forall j \in I \setminus \{i\}\}$.
\end{center}

\noindent That is, $B(\epsilon, I')$ stands for the set of schools each of which is the most preferred school for some student in $I'$. $T(\epsilon, O')$ stands for the set of students each of whom has the highest priority at some schools in $O'$.

A subset of students is identified as a \emph{mutual best group} in the following way. We first identify the set of schools ranked at the top place of these students' preferences. Then we identify the set of students pertaining the highest priorities at these schools. If the original subset of students is not only equivalent to the set of students identified just now but also the largest subset satisfying the above requirement, we call it a mutual best group. Formally,

\begin{Definition}
For each $\epsilon = (I, O,q, P, \succ)$, a subset of students $I' \subseteq I$ constitutes a \textbf{mutual best group} of $\epsilon$, denoted $G(\epsilon)$, if $I'$ is the largest subset of students satisfying the following condition:
\begin{center}

$T[\epsilon, B(\epsilon, I')]= I'$
\end{center}
\end{Definition}

Given a problem $\epsilon$, its mutual best group $G(\epsilon)$, and $B(\epsilon, G(\epsilon))$ which stands for the set of favorite schools of students in $G(\epsilon)$, if we remove $G(\epsilon)$ and $B(\epsilon, G(\epsilon))$ from the original problem $\epsilon$, we get a \textbf{subproblem} $\epsilon'=(I', O', P', q', \succ')$. In the subproblem, the set of students is $I'=I\backslash G(\epsilon)$, the quota vector is $q'=(q'_a)_{a\in O'}$ where $q'_a=q_a-|\{i\in G(\epsilon): aP_{i}b, \forall b \in O  \setminus \{a\}\}|$, the set of schools $O'$ is the set of schools with positive capacities after the removal of $G(\epsilon)$ and $B(\epsilon, G(\epsilon))$. The preference profile $P'$ is obtained by projection of $P$ on $I'$ and $O'$. Likewise, the priority profile $\succ'$ is obtained by projection of $\succ$ on $I'$ and $O'$. From the subproblem $\epsilon'$, we can also define the mutual best group and a new subproblem.

Therefore, given a problem $\epsilon$, we can define a sequence of subproblems and mutual best groups, which constitute a partition of students. Denote the \textbf{sequence of mutual best groups} by $MBG(\epsilon)=(I_1, I_2, \ldots, I_K)$, where $I_1$ is the mutual best group of the original problem ($I_1=G(\epsilon)$), $I_2$ is the mutual best group of the subproblem by removing $I_1$ and $B(\epsilon, I_1)$ from the original problem, and so on. Let $\epsilon=\epsilon_1$. We can also define a sequence of subproblems $(\epsilon_1, \epsilon_2, \ldots, \epsilon_K)$ such that for each $k\in \{1,2, \ldots, K\}$, $I_k$ is the mutual best group of the subproblem $\epsilon_k$, i.e., $I_k=G(\epsilon_k)$.


$~$

\subsection{Finding the mutual best group}

For any school choice problem $\epsilon$, the mutual best group $G(\epsilon)$ can be found via the following procedure:
\newline
\newline
\noindent \textbf{Finding the mutual best group}: Each student points to his favorite school and each school points to the student who has the highest priority for it. There is at least one cycle. The set of all students being in some cycle forms the mutual best group $G(\epsilon)$.

We illustrate the definition of mutual best group and how to find it by the following example.

\noindent \textbf{Example 1.} Let $I=\{1,2,3,4,5\}$, $O=\{a,b,c\}$ and $q_a=2, q_b=2, q_c = 1$. The preference profile $P$ and priority profile $\succ$ are given as follows:
\begin{center}
\begin{tabular}{cccccccccc}
\hline
$P_{1}$  & $P_{2}$  & $P_{3}$  & $P_{4}$  & $P_{5}$   &  & $\succ_{a}$ & $\succ_{b}$ & $\succ_{c}$ \\
\hline
$b$      &   $a$    & $a$  &   $c$    & $b$  &   & $1$ & $3$ & $4$ \\
$c$  &         $\varnothing$   &  $b$ &   $a$    & $a$   & &$4$ & $2$  & $5$ \\
$a$  &         $~$   &  $\varnothing$ &   $\varnothing$    & $c$   & &$5$ & $1$  & $3$ \\
$\varnothing$  & $~$&   $~$ & $~$&   $\varnothing$ & & $3$ & $5$ & $2$ \\
 \hline
\end{tabular}
\end{center}

Now let us find the sequence of mutual best group for this problem. Let each student point to his favorite school and each school point to the student who has the highest priority for it. There are two cycles:  $1\rightarrow b \rightarrow 3 \rightarrow a \rightarrow 1$ and $4\rightarrow c \rightarrow 4$. Therefore, the mutual best group is $G(\epsilon)=\{1,3, 4\}$.

If we remove $G(\epsilon)=\{1,3, 4\}$ and $B(\epsilon, G(\epsilon))$ from the original problem, we reach the following subproblem $\epsilon'=(I', O',q', P', \succ')$ where $I'=\{2, 5\}$, $O'=\{a, b\}$, $q_a^{'}=1$, and $q_b^{'}=1$. For the subproblem, let each student point to his favorite school and each school point to the student who has the highest priority for it. There is only one cycle of $2 \rightarrow a \rightarrow 5 \rightarrow b \rightarrow 2$. The mutual best group of the subproblem is $G(\epsilon')=\{2, 5\}$. Therefore, we partition the students into two mutual best groups $MBG(\epsilon)=(\{1,3, 4\}, \{2, 5\})$.

If more than one cycle form in a certain step, students in all of these cycles form a mutual best group.

\subsection{Redefining the TTCM}

Building on the content above, we can now define the (student-optimal) top trading cycles algorithm in a more compact form.
\begin{Definition}
Given $\epsilon$, and the sequence of mutual best groups $MBG(\epsilon)=\{I_1, I_2, \ldots, I_K\}$, the top trading cycles mechanism, denoted $\varphi^T$, is defined as follows:
\begin{center}
$\varphi^T_i(\epsilon)= B(\epsilon^{1}, i)$, $\forall i\in I_1$

$\vdots$

$\varphi^T_i(\epsilon)= B(\epsilon^{k}, i)$, $\forall i\in I_k$

$\vdots$

$\varphi^T_i(\epsilon)= B(\epsilon^{K}, i)$, $\forall i\in I_K$
\end{center}
\end{Definition}
The top trading cycles mechanism is essentially \lq\lq sequentially prioritized \rq\rq, with the order of students being determined endogenously and in a group manner. Given a problem and the sequence of mutual best groups determined by it, students in the first mutual best group get their favorite schools, students in the second mutual best group get their favorite schools among what remains, and so on.









\section{Two new axioms}
\subsection{A weaker stability axiom}

We now propose a new axiom which reflects fairness on priorities. It requires that if a students has the top quota priority for a school, then the student weakly prefers his own assignment to this school. This axiom can be considered as a generalization of respect of top priorities defined in \cite{abdulkadiroglu2010role}. Formally,

\begin{Definition}
A matching $\mu$ \textbf{respects quota-priorities} at a problem $\epsilon$, if for each $i\in I$ and each $a \in \widetilde{O}$,
\[
|\{j\in I: j\succ_a i\}|<q_a \Rightarrow \mu_i R_i a.
\]
\end{Definition}

Given a matching $\mu$ with respect to $\epsilon$, and a subset of students $I' \subset I$, let the subproblem by removing $I'$ and $\mu_{I'}$ be denoted $\epsilon^{\mu}_{-I'}$. We say that a matching satisfies quota-rationality if its projection in the subproblem by removing any subset of students and their assigned schools respects quota-priorities.

\begin{Definition}
A matching $\mu$ satisfies \textbf{quota-rationality} at a problem $\epsilon$ if for each (possibly empty) $I' \subset I$, $\mu_{-I'}$ respects quota-priorities at $\epsilon^{\mu}_{-I'}$.
\end{Definition}

Coincidentally, quota-rationality is equivalent to stability in school choice setting. The next proposition formalizes this idea and it can be considered as a generalization of \cite{abdulkadiroglu2010role}'s theorem 6.

\begin{Proposition}
A matching satisfies quota-rationality if and only if it is stable.
\end{Proposition}

\begin{proof}
The if part: If a matching $\mu$ is stable at a problem $\epsilon$, then evidently it respects quota-priorities. If we remove a subset $I'\subset I$ and their assignment $\mu_{I'}$ from the problem, then we have $\mu_{-I'}$ is stable at $\epsilon^{\mu}_{-I'}$, and this further implies that $\mu_{-I'}$ respects quota-priorities at $\epsilon^{\mu}_{-I'}$ for each $I'\subset I$, which is exactly the definition of quota-rationality.

The only if part: suppose by contradiction that a match $\mu$ satisfies quota-rationality, but is not stable at a problem $\epsilon$. There are two possible cases. \\
Case 1. There exist a student $i$ and a school $a$ such that $a P_{i} \mu_{i}$ and $|\mu_{a}| < q_{a}$.

Let $I'$ be the set of students with higher priorities than $i$ at $s$. By removing $I'$ and $\mu_{I'}$, we obtain a subproblem $\epsilon^{\mu}_{-I'}$. Since $i$ has the highest priority at $a$ for the subproblem, by quota-rationality, $\mu_{i} R_{i} a$. It contradicts with the initial supposition of Case 1. \\
Case 2. There exist a student $i$ and a school $a$ such that $a P_{i} \mu_{i}$ and there is some $j \in \mu_{a}$ with $i \succ_{a} j$.

Let $I' = I \setminus \{i, j\}$. By removing $I'$ and $\mu_{I'}$, we obtain a subproblem $\epsilon^{\mu}_{-I'}$. Since $i i\succ_{a} j$ and $\mu_{j} = a$, by quota-rationality, $\mu_{i} R_{i} a$. It contradicts with the initial supposition of Case~2.
\end{proof}

It is easy to construct examples showing that TTCM violates quota-rationality. Next we propose a weaker version of quota-rationality which is satisfied by TTCM. A matching satisfies MBG-quota-rationality if its projection in any subproblem by removing any subset of mutual best groups and their assigned schools respects quota-rationalities.

\begin{Definition}
A matching $\mu$ satisfies MBG-quota-rationality at a problem $\epsilon$ if for each $I'\in 2^{MBG(\epsilon)}$, $\mu_{-I'}$ respects quota-priorities at $\epsilon^{\mu}_{-I'}$.\footnote{Here $2^{MBG(\epsilon)}$ stands for the power set of $MBG(\epsilon)$, which is the set of all subsets of it.} A mechanism satisfies \textbf{MBG-quota-rationality} if it selects a matching satisfying MBG-quota-rationality for each problem.
\end{Definition}

It follows immediately from the definitions and Proposition 1 that stability implies MBG-quota-rationality. Since SOSM and TTCM both satisfy MBG-quota-rationality, we could interpret that in some sense the axiom reflects the similarities in fairness terms between these two mechanisms.

\subsection{A stronger efficiency axiom}

We propose another axiom  with a similar idea as robust stability studied in \cite{kojima2011robust}, except that our focus is on the efficiency side. First, we introduce a few related definitions.

Recall that given a matching $\mu$ and $I'\subseteq I$, $M(I', \mu_{I'})$ is the set of all possible matchings between $I'$ and $\mu_{I'}$, and $m(I', \mu_{I'})$ is a generic element of $M(I', \mu_{I'})$. We first introduce an axiom related to joint manipulations by a group of students. This axiom requires that no group of students can weakly benefit by misreporting their preferences or first misreporting preferences and then reallocate their assignments.

\begin{Definition}
A mechanism $\varphi$ is \textbf{collusion-proof} if for each problem $P$, each nonempty $I' \subseteq I$, and each $P_{I'}^{'}$, there exists no $m[I',\varphi_{I'}(P_{I'}^{'}, P_{-I'})] \in M[I',\varphi_{I'}(P_{I'}^{'}, P_{-I'})]$ such that
\begin{itemize}
\item[(i)]
$m_i[I',\varphi_{I'}(P_{I'}^{'}, P_{-I'})] R_i \varphi_i(P)$ for all $i\in I'$,  and
\item[(ii)]
$m_j[I',\varphi_{I'}(P_{I'}^{'}, P_{-I'})] P_j \varphi_j(P)$ for some $j \in I'$.
\end{itemize}
\end{Definition}

Collusion-proofness is a demanding axiom for school choice problems. It implies reallocation-proofness,


We next define a weaker version of collusion-proofness which is satisfied by TTCM.
\begin{Definition}
A mechanism $\varphi$ is \textbf{MBG-collusion-proof} if for each problem $P$, each nonempty $I' \in MBG(\epsilon)$, and each $P_{I'}^{'}$, there exists no $m[I',\varphi_{I'}(P_{I'}^{'}, P_{-I'})] \in M[I',\varphi_{I'}(P_{I'}^{'}, P_{-I'})]$ such that
\begin{itemize}
\item[(i)]
$m_i[I',\varphi_{I'}(P_{I'}^{'}, P_{-I'})] R_i \varphi_i(P)$ for all $i\in I'$,  and
\item[(ii)]
$m_j[I',\varphi_{I'}(P_{I'}^{'}, P_{-I'})] P_j \varphi_j(P)$ for some $j\in I'$.
\end{itemize}
\end{Definition}

A mechanism satisfies robust efficiency if (1) it is Pareto efficient; and (2) it is collusion-proof, i.e., no subset of students can gain (all members are weakly better off while some are strictly better off) by either reporting a false preference profile or first reporting a false preference profile and then reallocate their assigned schools within the subset. Formally,
\begin{Definition}
A mechanism $\varphi$ is \textbf{robustly efficient} if,
\begin{enumerate}
\renewcommand{\labelenumi}{(\theenumi)}
\item it is \textbf{Pareto efficient};
\item it is \textbf{collusion-proof}.
\end{enumerate}
\end{Definition}

Now we are ready to define our second main axiom. It requires that for each problem, the matching selected is Pareto efficient, and moreover, there exists no mutual best group that can benefit from first jointly misreporting their preferences and then execute a Pareto improvement within the group. Formally,

\begin{Definition}
A mechanism $\varphi$ is \textbf{MBG-robustly efficient} if
\begin{enumerate}
\renewcommand{\labelenumi}{(\theenumi)}
  \item $\varphi$ is \textbf{Pareto efficient};
  \item $\varphi$ is \textbf{MBG-collusion proof}.
\end{enumerate}
\end{Definition}

The following proposition shows the logical relationship between reallocation-proofness and MBG-robust efficiency (MBG-collusion-proofness).
\begin{Proposition}
MBG-robust efficiency (MBG-collusion-proofness) is independent with reallocation-proofness.
\end{Proposition}

We show the \textbf{independence of axioms} below by examples. In Example 3, we present a mechanism which satisfies MBG-collusion-proofness but violates reallocation-proofness. The serial dictatorship mechanism (SD) satisfies reallocation-proofness but violates MBG-collusion-proofness, which is illustrated in Example 4.

\textbf{Example 3} Let $I=\{1,2,3\}$, $O=\{a,b\}$ and $q_a=1, q_b=2$. The preference profile $P$ and priority profile $\succ$ are given as follows:

\begin{center}
\begin{tabular}{cccccccccc}
\hline
$P_{1}$  & $P_{2}$  & $P_{3}$  & $P_{1}^{'}$ &   $P_{2}^{'} $   &  & $\succ_{a}$ & $\succ_{b}$ \\
\hline
$b$      &   $a$    & $a$  & $a$& $b$&   & $1$ & $3$ \\
$a$  &         $b$   &  $b$ & $b$& $a$ & &$2$ & $1$ \\
$\varnothing$       & $\varnothing$&   $\varnothing$ &$\varnothing$&$\varnothing$& & $3$ & $2$ \\
 \hline
\end{tabular}
\end{center}

\noindent Let $\psi$ be a mechanism such that

\[
\psi(P)=
\begin{cases}
(a,b,\varnothing), \mbox{ if~~} P=(P_{1}^{'}, P_{2}^{'}, P_3).  \\
(\varnothing,\varnothing,\varnothing), \mbox{ otherwise.}\\
  \end{cases}
\]

\noindent Note that the sequence of mutual best groups of this problem is $MBG(\epsilon)=\{\{1,3\},\{2\}\}$. It is straightforward to verify that $\psi$ satisfies MBG-collusion-proofness but violates reallocation-proofness.

$~$

TTCM satisfies MBG-robust efficiency, while SOSM and SD violates it. Example 4 below presents this result compactly.

\textbf{Example 4 (Example 3 revisits.)} Let $I=\{1,2,3\}$, $O=\{a,b\}$ and $q_a=1, q_b=2$. The preference profile $P$ and priority profile $\succ$ are given as follows:

\begin{center}
\begin{tabular}{cccccccccc}
\hline
$P_{1}$  & $P_{2}$  & $P_{3}$  & $P_{1}^{'}$ &   $P_{3}^{'} $   &  & $\succ_{a}$ & $\succ_{b}$ \\
\hline
$b$      &   $a$    & $a$  & $a$& $b$&   & $1$ & $3$ \\
$a$  &         $b$   &  $b$ & $b$& $a$ & &$2$ & $1$ \\
$\varnothing$       & $\varnothing$&   $\varnothing$ &$\varnothing$&$\varnothing$& & $3$ & $2$ \\
 \hline
\end{tabular}
\end{center}

Let $P=(P_{1},P_{2},P_{3})$ and $P^{'}=(P_{1}^{'},P_{2},P_{3}^{'})$. The sequence of mutual best groups of this problem is $MBG(\epsilon)=\{\{1,3\},\{2\}\}$.

\begin{center}
\begin{tabular}{cccccccccc}
\hline
&Mechanism & Matching     & & 1& 2  & 3 & MBG-robust efficiency\\
\hline
1&TTCM  &   $\varphi^T(P)$     & & $b$& $b$& $a$ &\\
2&TTCM  &   $\varphi^T(P^{'})$ & & $a$& $b$& $b$&\\
3&TTCM  &   Reallocation  &      & $b$& $b$& $a$& Yes\\
4&SOSM  &   $\varphi^S(P)$     & & $b$& $a$& $b$&\\
5&SOSM  &   $\varphi^S(P^{'})$ & & $a$& $b$& $b$&\\
6&SOSM  &   Reallocation & & $b$& $b$& $a$& No\\
7&SD ($f=1,2,3$) &  $\varphi^f(P)$      & &$b$& $a$& $b$&\\
8&SD ($f=1,2,3$) &  $\varphi^f(P^{'})$  & &$a$& $b$& $b$&\\
9&SD ($f=1,2,3$) &  Reallocation  & &$b$& $b$& $a$& No\\
 \hline
\end{tabular}
\end{center}

Under $\varphi^T$, if students $\{1,3\}$ report a false preference profile of $(P'_1, P'_3)$, through comparing the matchings of the 1st and 2nd rows we know that neither of them become strictly better off, and if they exchange (reallocate) their assigned schools, through comparing the matchings of the 2nd and 3rd rows we know that neither of them become strictly better off. Actually, there exists no space for students $\{1,3\}$ to gain no matter what preferences they report, i.e., \textbf{TTCM satisfies MBG-robust efficiency}.

Under $\varphi^S$, if students $\{1,3\}$ report a false preference profile of $(P'_1, P'_3)$, through comparing the matchings of the 4th and 5th rows we know that neither of them become strictly better off. The general reason for this is because the priority structure is Ergin-acyclic and SOSM recovers group strategy-proofness. But if they exchange (reallocate) their assigned schools, through comparing the matchings of the 5th and 6th rows we know that student 3 become strictly better off (His assignment changes from $b$ to $a$.), while student 1's assignment remain the same as the no misreporting case. The above statement shows that \textbf{SOSM violates MBG-robust efficiency even under Ergin-acyclic priority structures.}

Under $\varphi^f$, if students $\{1,3\}$ report a false preference profile of $(P'_1, P'_3)$, through comparing the matchings of the 7th and 8th rows we know that neither of them become strictly better off. But if they exchange (reallocate) their assigned schools, through comparing the matchings of the 8th and 9th rows we know that student 3 become strictly better off (His assignment changes from $b$ to $a$.), while student 1's assignment remain the same as the no misreporting case. The above statement shows that \textbf{SD violates MBG-robust efficiency.}

\section{Characterizations}

Our main result characterizes the TTCM by imposing MBG-collusion-proofness, together with efficiency and MBG-quota-rationailty.

\begin{Theorem}
A mechanism $\varphi$ satisfies MBG-quota-rationality and MBG-robust efficiency if and only if $\varphi=\varphi^T$.
\end{Theorem}

\begin{proof}
The if part is straightforward.  We prove the only if part through one definition and two lemmas.

We next define a new axiom based on MBGs. A matching respects MBGs if each mutual best group in the sequence is assigned the set of favorite schools of students in this mutual best group. Formally,

\begin{Definition}
A mechanism $\varphi$ \textbf{respects mutual best groups(MBGs)} if for each $\epsilon$ and $MBG(\epsilon)=\{I_1, I_2, \ldots, I_k, \ldots, I_K\}$,
\begin{center}
$\cup_{i \in _{I_1}}\varphi_{i}(\epsilon)= B(\epsilon^{1}, I_1)$

$\vdots$

$\cup_{i \in _{I_k}}\varphi_{i}(\epsilon)= B(\epsilon^{k}, I_k)$

$\vdots$

$\cup_{i \in _{I_K}}\varphi_{i}(\epsilon)= B(\epsilon^{K}, I_K)$

\end{center}
\end{Definition}

\begin{Lemma}
If a mechanism $\varphi$ satisfies MBG-quota-rationality and MBG-robust efficiency, then $\varphi$ respects mutual best groups.
\end{Lemma}
\begin{proof}
We proceed by contradiction. Suppose a mechanism satisfies MBG-quota-rationality and MBG-robust efficiency but does not respect mutual best groups. Without loss of generality, suppose $\cup_{i \in _{I_1}}\varphi_{I_1}(\epsilon)\neq B(\epsilon, I_1)$. This implies that there exists $i\in I_1$ such that $i$ is not assigned his favorite school under $\varphi(\epsilon)$, i.e., $\varphi_i(\epsilon)\neq B(\epsilon, i)$.

Now, if each $i\in I_1$ reports the preference $P'_i: \{a| a\in B(\epsilon, I_1)\&T(\epsilon, a)=i\} P'_i \varnothing$, then by MBG-quota-rationality of $\varphi$, $\varphi_{I_1}(P_{I_1}^{'}, P_{-I_1})= B(\epsilon, I_1)$. Furthermore, through reallocating $B(\epsilon, I_1)$ within the group $I_1$, all members can possibly get their favorite schools, which makes all members weakly better off, while at least one member strictly better off. Therefore, through misreporting and reallocating, the group achieves Pareto improvement, which contradicts MBG-robust efficiency of $\varphi$.
\end{proof}

\begin{Lemma}
If a mechanism $\varphi$ respects mutual best groups and is Pareto efficient, then $\varphi=\varphi^T$.
\end{Lemma}
\begin{proof}
Let $\varphi(\epsilon)$ and $\varphi^T(\epsilon)$ be $\mu$ and $\mu^T$, respectively. By the definition of mutual best groups and top trading cycles algorithm, we know that once $\mu\neq\mu^T$, $\mu^T$ Pareto dominates $\mu$, which contradicts Pareto efficiency of $\mu$.
\end{proof}

Lemma 1 and 2 complete the proof of the theorem.
\end{proof}

$~$

\textbf{Independence of axioms:} The SOSM satisfies MBG-quota-rationality, but violates MBG-robust efficiency. The mechanism $\tilde{\varphi}$ defined below satisfies MBG-robust efficiency but violates MBG-quota-rationality.

$~$

\begin{Example}
Let $\mathcal{E}'\subset \mathcal{E}$. A problem $(I, O, P, \succ, q)=\epsilon \in \mathcal{E}'$ if for each $i,j\in I$, $P_i=P_j$. Given an ordering of agents $f$, let $SD^f$ be the serial dictatorship induced by $f$.

Let the mechanism $\tilde{\varphi}$ be defined as follows: for each problem $\epsilon \in \mathcal{E}$,
\begin{itemize}
\item[(i)]
$\tilde{\varphi}(\epsilon) =SD^{f}(\epsilon)$, if $\epsilon \in \mathcal{E}'$;
\item[(ii)]
$\tilde{\varphi}_{i}(\epsilon) = \varphi^{T}(\epsilon)$, otherwise.
\end{itemize}
\end{Example}

$~$

\textbf{Remark 1:} As shown in \cite{ergin2002efficient}, SOSM is Pareto efficient if and only if a priority structure is Ergin-acyclic. Even if we restrict our attention to this smaller domain of priority structures, it turns out that TTCM is still the only mechanism satisfying MBG-quota-rationality and MBG-robust efficiency. This result reveals that immunity to collusion within mutual best groups (MBG-collusion-proofness) is the key difference between TTCM and SOSM.

\textbf{Remark 2:} In the model where each school has only one seat, \cite{abdulkadiroglu2010role} characterize the TTCM by strategy-proofness, efficiency, and recursive respect of top priorities. Since MBG-quota-rationality can be considered as a natural generalization of recursive respect of top priorities to the multiple seats model, the question then arises whether the TTCM is the unique mechanism satisfying MBG-quota-rationality, strategy-proofness, and efficiency.

It turns out that there are other mechanisms satisfying these three axioms. We present one such mechanism.

\begin{Definition}
Let the mechanism $\hat{\varphi}$ be defined as follows: for each problem $\epsilon \in \mathcal{E}$,
\begin{itemize}
\item[(i)]
$\hat{\varphi}(\epsilon) = \varphi^{S}(\epsilon)$, if $\succ$ is Ergin-acyclic;
\item[(ii)]
$\hat{\varphi}(\epsilon) = \varphi^{T}(\epsilon)$, otherwise.
\end{itemize}
\end{Definition}

This new mechanism differs from the TTCM only in a small subdomain. It is straightforward to verify that $\hat{\varphi}$ satisfies MBG-quota-rationality, strategy-proofness, and efficiency. Note that this mechanism even satisfies group strategy-proofness. It shows that the characterization result of \cite{abdulkadiroglu2010role} in one-seat model can not be extended easily to the general model.

\section{Concluding remarks}

Our main result characterizes the TTCM when each school may have more than one seats, by MBG-quota-rationality and MBG-robust efficiency. This constructs a new basis to compare TTCM with the other school choice mechanisms like BOSM, SOSM, SD, EADAM (\cite{kesten2010school}, \cite{tang2014new}), and so on.

We propose a new axiom, MBG-collusion-proofness. When the priority structure is Ergin-acyclic, SOSM recovers Pareto efficiency, group strategy-proofness and consistency\footnote{\cite{kesten2006two} shows that TTCM violates consistency under Ergin-acyclic priority structures.}. It seems that SOSM dominates TTCM under such circumstances. However, our work shows otherwise. If ex post reallocation of schools is not forbidden, TTCM has brighter prospects than SOSM because it satisfies MBG-robust efficiency, i.e., it can prevent at least students in each mutual best group from colluding.

One possible direction for future research is to further investigate robust efficiency first defined in this paper in various theoretical environments.
\nocite{2018endowments}
\nocite{moulin1995Cooperative}

\bibliography{TTCM}

\begin{thebibliography}{24}
\providecommand{\natexlab}[1]{#1}

\bibitem[{Abdulkadiro{\u{g}}lu \textit{et~al.}(2020)Abdulkadiro{\u{g}}lu, Che,
  Pathak, Roth and Tercieux}]{abdulkadirouglu2020efficiency}
\textsc{Abdulkadiro{\u{g}}lu, A.}, \textsc{Che, Y.-K.}, \textsc{Pathak, P.~A.},
  \textsc{Roth, A.~E.} and \textsc{Tercieux, O.} (2020). Efficiency, justified
  envy, and incentives in priority-based matching. \textit{American Economic
  Review: Insights}, \textbf{2}~(4), 425--42.

\bibitem[{Abdulkadiro{\u{g}}lu and
  S{\"o}nmez(2003)}]{abdulkadirouglu2003school}
\textsc{---} and \textsc{S{\"o}nmez, T.} (2003). School choice: a mechanism
  design approach. \textit{American Economic Review}, \textbf{93}~(3),
  729--747.

\bibitem[{Abdulkadiro\u{g}lu and Che(2010)}]{abdulkadiroglu2010role}
\textsc{Abdulkadiro\u{g}lu, A.} and \textsc{Che, Y.} (2010). {The role of
  priorities in assigning indivisible objects: a Characterization of top
  trading cycles}.

\bibitem[{Afacan(2012)}]{afacan2012Group}
\textsc{Afacan, O.~M.} (2012). Group robust stability in matching markets.
  \textit{Games and Economic Behavior}, \textbf{74}, 394--398.

\bibitem[{Afacan(2013)}]{afacan2013alternative}
\textsc{---} (2013). {Alternative characterizations of Boston mechanism}.
  \textit{Mathematical Social Sciences, \emph{forthcoming}}.

\bibitem[{Chen and Heo(2021)}]{chen2021acyclic}
\textsc{Chen, S.} and \textsc{Heo, E.~J.} (2021). Acyclic priority profiles in
  school choice: Characterizations. \textit{Journal of Mathematical Economics},
  \textbf{92}, 22--30.

\bibitem[{Chen(2016)}]{chen2016new}
\textsc{Chen, Y.} (2016). New axioms for immediate acceptance. \textit{Review
  of Economic Design}, \textbf{20}~(4), 329--337.

\bibitem[{Chen(2017)}]{chen2017new}
\textsc{---} (2017). New axioms for deferred acceptance. \textit{Social Choice
  and Welfare}, \textbf{48}~(2), 393--408.

\bibitem[{Ehlers and Klaus(2014)}]{ehlers2014strategy}
\textsc{Ehlers, L.} and \textsc{Klaus, B.} (2014). Strategy-proofness makes the
  difference: deferred-acceptance with responsive priorities.
  \textit{Mathematics of Operations Research}, \textbf{39}~(4), 949--966.

\bibitem[{Ergin(2002)}]{ergin2002efficient}
\textsc{Ergin, H.} (2002). Efficient resource allocation on the basis of
  priorities. \textit{Econometrica}, \textbf{70}~(6), 2489--2497.

\bibitem[{Fujinaka and Wakayama(2018)}]{2018endowments}
\textsc{Fujinaka, Y.} and \textsc{Wakayama, T.} (2018).
  Endowments-swapping-proof house allocation. \textit{Games and Economic
  Behavior}, \textbf{111}, 187--202.

\bibitem[{Gale and Shapley(1962)}]{gale1962college}
\textsc{Gale, D.} and \textsc{Shapley, L.} (1962). College admissions and the
  stability of marriage. \textit{The American Mathematical Monthly},
  \textbf{69}~(1), 9--15.

\bibitem[{Haeringer and Klijn(2009)}]{Haeringer2009Constrained}
\textsc{Haeringer, G.} and \textsc{Klijn, F.} (2009). Constrained school
  choice. \textit{Journal of Economic Theory}, \textbf{144}~(5), 1921--1947.

\bibitem[{Kesten(2006)}]{kesten2006two}
\textsc{Kesten, O.} (2006). On two competing mechanisms for priority-based
  allocation problems. \textit{Journal of Economic Theory}, \textbf{127}~(1),
  155--171.

\bibitem[{Kesten(2010)}]{kesten2010school}
\textsc{---} (2010). School choice with consent. \textit{The Quarterly Journal
  of Economics}, \textbf{125}~(3), 1297--1348.

\bibitem[{Kojima(2011)}]{kojima2011robust}
\textsc{Kojima, F.} (2011). Robust stability in matching markets.
  \textit{Theoretical Economics}, \textbf{6}~(2), 257--267.

\bibitem[{Morrill(2011)}]{morrill2011alternative}
\textsc{Morrill, T.} (2011). An alternative characterization of top trading
  cycles. \textit{Economic Theory}.

\bibitem[{Morrill(2013)}]{morrill2013alternative}
\textsc{---} (2013). An alternative characterization of the deferred acceptance
  algorithm. \textit{International Journal of Game Theory}, \textbf{42}~(1),
  19--28.

\bibitem[{Moulin(1995)}]{moulin1995Cooperative}
\textsc{Moulin, H.} (1995). \textit{Cooperative Microeconomics}. Princeton
  University Press.

\bibitem[{P{\'a}pai(2000)}]{papai2000strategyproof}
\textsc{P{\'a}pai, S.} (2000). Strategyproof assignment by hierarchical
  exchange. \textit{Econometrica}, \textbf{68}~(6), 1403--1433.

\bibitem[{Pycia and {\"U}nver(2017)}]{pycia2017incentive}
\textsc{Pycia, M.} and \textsc{{\"U}nver, M.~U.} (2017). Incentive compatible
  allocation and exchange of discrete resources. \textit{Theoretical
  Economics}, \textbf{12}~(1), 287--329.

\bibitem[{Rong \textit{et~al.}(2020)Rong, Tang and Zhang}]{rong2020stable}
\textsc{Rong, K.}, \textsc{Tang, Q.} and \textsc{Zhang, Y.} (2020). On stable
  and efficient mechanisms for priority-based allocation problems.
  \textit{Journal of Economic Theory}, \textbf{187}, 105014.

\bibitem[{Shapley and Scarf(1974)}]{shapley1974cores}
\textsc{Shapley, L.} and \textsc{Scarf, H.} (1974). On cores and
  indivisibility. \textit{Journal of mathematical economics}, \textbf{1}~(1),
  23--37.

\bibitem[{Tang and Yu(2014)}]{tang2014new}
\textsc{Tang, Q.} and \textsc{Yu, J.} (2014). A new perspective on kesten's
  school choice with consent idea. \textit{Journal of Economic Theory},
  \textbf{154}, 543--561.

\end{thebibliography}
\bibliographystyle{ecca}
\end{document}